\documentclass[%
 reprint,longbibliography,preprintnumbers,
nofootinbib,
 amsmath,amssymb,
 aps,
prb,
]{revtex4-2}
\pdfoutput=1
\usepackage[utf8]{inputenc}
\usepackage{flushend}
\usepackage{dcolumn}
\usepackage{bm}
\usepackage{balance}

\usepackage[normalem]{ulem}

\usepackage[colorlinks = true,
            linkcolor = blue,
            urlcolor  = blue,
            citecolor =green,
            anchorcolor = blue]{hyperref}
\usepackage{verbatim}
\usepackage{color,ulem}
\usepackage[english]{babel}

\usepackage[utf8]{inputenc}
\newcommand*\initfamily{\usefont{U}{Starburst}{xl}{n}}\initfamily

\newcommand{\beq}{\begin{eqnarray}}
\newcommand{\eeq}{\end{eqnarray}}
\usepackage{amsmath}
\usepackage{tikz}
\usetikzlibrary{decorations.pathmorphing}
\usetikzlibrary{shapes.misc}
\tikzset{cross/.style={cross out, draw=black, minimum size=8*(#1-\pgflinewidth), inner sep=0pt, outer sep=0pt},
cross/.default={1pt}}
\usetikzlibrary{patterns,math}
\begin{document}

\title{Quantum confinement theory of ultra-thin films: electronic, thermal and superconducting properties}

\author{\textbf{Alessio Zaccone}$^{1}$}%
 \email{alessio.zaccone@unimi.it}
 
 \vspace{1cm}
 
\affiliation{$^{1}$Department of Physics ``A. Pontremoli'', University of Milan, via Celoria 16,
20133 Milan, Italy.}

\begin{abstract}
The miniaturization of electronic devices has led to the prominence, in technological applications, of ultra-thin films with a thickness ranging from a few tens of nanometers to just about 1-2 nanometers. 
While these materials are still effectively 3D in many respects, traditional theories as well as ab initio methods struggle to describe their properties as measured in experiments.  
In particular, standard approaches to quantum confinement rely on hard-wall boundary conditions, which neglect the unavoidable, ubiquitous, atomic-scale irregularities of the interface. Recently, a unified theoretical approach to quantum confinement has been proposed which is able to effectively take the real nature of the interface into account, and can efficiently be implemented in synergy with microscopic theories. Its predictions for the electronic properties such as electrical conductivity of semiconductor thin films or critical temperature of superconducting thin films, have been successfully verified in comparison with experimental data. The same confinement principles lead to new laws for the phonon density of states and for the heat capacity of thin films, again in agreement with the available experimental data. 
\end{abstract}

\maketitle
\section{Introduction}
The physical properties of thin films are vital for many technological applications, ranging from optical mirrors to solar cells \cite{solar_science,Friend}, and they are also of interest for fundamental condensed matter physics. 
For example, our everyday life would be unthinkable without the achievements of microelectronics, a revolution that began with the discovery of transistors \cite{Bardeen}. Ever since, the main strategy to make more powerful electronic devices is to shrink the size of semiconductor blocks in a microchip, with the newest types of MOSFET reaching sizes in the range between 7 nm and 22 nm. 
A different strategy to push the boundaries of computing power is provided by quantum computers. These are based on qubits, which are physically realized by ultra-thin superconducting elements in the thickness range from few tens of nanometers to about 100 nm. Thin films of superconducting materials are also promising as an alternative route to microelectronics, thanks to the recently discovered supercurrent field effect, where electric fields are used to suppress the supercurrent in ultra-thin films \cite{Giazotto1,Giazotto2}. 
Metallic and superconducting ultra-thin films are also of great technological importance for their thermal properties, in particular for their application as single-photon detectors and, again, as components for superconducting nanoelectronics. In all these applications, it is crucial to effectively control the heat removal and the heat dissipation at cryogenic temperatures, which ultimately means tackling phonon transport problems under nanoscale confinement. 

At the very heart of the problem of understanding superconductivity, or electronic and heat transport, in nanometric thin films, lies the fundamental problem of effectively describing the propagation of wavefunctions in nano-confinement. The most difficult problem is posed by the presence of an interface which impedes or limits the propagation of the wavefunction across it. This is because of the irregularity of the interface, which is unavoidable in any experimentally realized film. Even for the most regular surface of a perfectly crystalline film with no defects, assuming that the wavefunction vanishes exactly at the same coordinate of the surface is a strong idealization, made unphysical by the atomic roughness of the surface.

A new approach, originated from the study of elasticity of thin liquid films \cite{PNAS2020}, has recently been investigated, which can manage the description of wavefunctions under nano-confinement without having to assume a fixed hard-wall boundary condition.

Furthermore, this theory provides a simple way of obtaining analytical closed-form expressions for key physical quantities, such as the electronic density of states (DOS), the phonon DOS, the Fermi energy, as a function of the film thickness. This allows the theory to be incorporated into microscopic frameworks and models of electronic and vibrational properties. Agreement with experimentally observed quantities has been achieved over the past few years for different quantities (electrical conductivity, superconducting critical temperature, specific heat) and for diverse materials, ranging from semiconductors to metals to insulators. Particularly striking has been the prediction of the superconducting critical temperature as a function of nanometric film thickness for two real materials, aluminum and lead, with no adjustable parameters, by implementing this quantum confinement strategy within the Eliashberg theory of superconductivity \cite{Ummarino_2025}.

In this Perspective paper we attempt to provide a succinct pedagogical introduction to these latest developments. We will consistently focus on thin films that remain 3D even in the quasi-2D or ultra-thin (sub-nanometer) limit. This treatment leaves the perfect 2D limit, i.e. the atomic monolayer, out of the current discussion. This choice is motivated by the fact that the basic physical laws change discontinuously upon going from a 3D multi-layer to a perfectly 2D monolayer, as reflect e.g. in different functional forms of the density of states for both phonons and electrons and in many other properties and phenomena, the mechanism of which radically changes upon taking the perfect monolayer limit. For a discussion of these issues, cf. Ref. \cite{Travaglino_2022}.

\section{Quantum confinement of wavefunctions in real materials}
\begin{figure}[h]
\centering
\includegraphics[width=\linewidth]{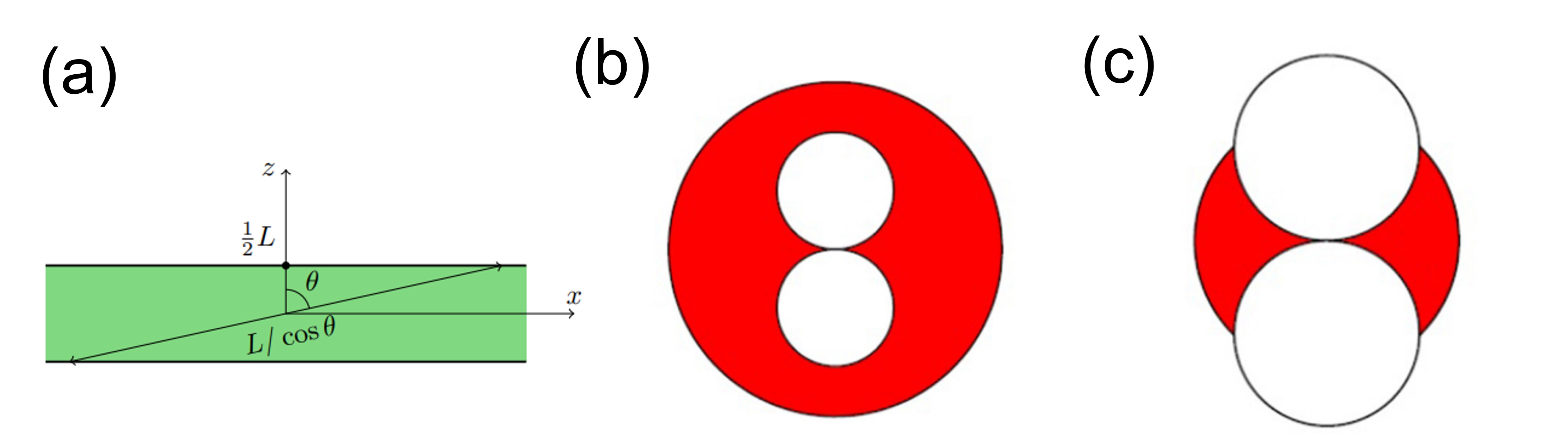}
\caption{Panel (a) shows the thin film geometry in real space (confined along the $z$-axis but unconfined along the $x$ and $y$ axis), with the maximum wavelength of a free carrier that corresponds to a certain polar angle $\theta$. Panels (b)-(c) shows the corresponding geometry of occupied quasiparticle states in $k$-space. For free electrons, the outer Fermi sphere (of radius $k_{F}$) contains two symmetric spheres of hole pockets (states suppressed by confinement), i.e. states in $k$-space that remain unoccupied due to confinement along the $z$-axis. In (b), for weak confinement, the two hole pockets are well within the Fermi surface, which remains spherical. In (c), for strong confinement (e.g. quasi-2D films), the hole pockets of forbidden states have grown to the point that the Fermi surface gets significantly distorted into a surface belonging to a different homotopy group $\mathbb{Z}$. See Refs. \cite{Travaglino_2022,Travaglino_2023,Phillips} for a detailed mathematical derivation of these results. }
\label{fig1}
\end{figure}

We schematize the thin film as a 3D material with confinement along the vertical $z$ direction, and consider it unconfined along the two other Cartesian directions, i.e. in the $xy$ plane, as schematically depicted in Fig. \ref{fig1} for the generic case of non-interacting quasiparticles (e.g. phonons or electrons). For electrons, the red sphere is the Fermi sphere, while for phonons it is the Debye sphere.

As derived in Ref. \cite{Phillips,Travaglino_2022, Travaglino_2023,Yu_2022}, along the direction $\theta$ (cfr. Fig. \ref{fig1}(a)), plane-wave quasiparticles with wavelength 
\begin{equation}
    \lambda >\lambda_{max} = \frac{L}{\cos\theta} \label{cutoff}
\end{equation}
cannot propagate in the thin film. Here, the polar angle $\theta$ of the propagation direction is measured with respect to the vertical $z$ axis (cfr. Fig. \ref{fig1}(a)).

As stated in all quantum mechanics textbooks, the momenta $k_{x}$, $k_{y}$, $k_{z}$ of a quantum particle in a (small) box are, in general, discretized whenever standard vanishing (or hard-wall) boundary conditions (BCs) are chosen for the governing Schr{\"o}dinger equation. In particular, by imposing that the quasiparticle wavefunction $\psi$ vanishes exactly at the borders of the, e.g., rectangular, box, one obtains plane-wave forms of the type $\psi \sim \sin(k_{x} x)\sin(k_{y} y) \sin(k_{z} z)$. The hard-wall BCs lead to a discrete set of values, $k_z =\pi n/L$, with $n$ an integer number.
Accordingly, one expects the minimum wavevector in the system to be given by:
\begin{equation}
k_{z,min} = \pi/L.
\end{equation}
However, the assumption that the minimum value of $k_{z}$ is equal to $\pi/L$ is valid only when the standard hard-wall BCs are strictly enforced at the boundaries of the rectangular box. 

\begin{figure}[ht!]
    \centering
    \includegraphics[width=\linewidth]{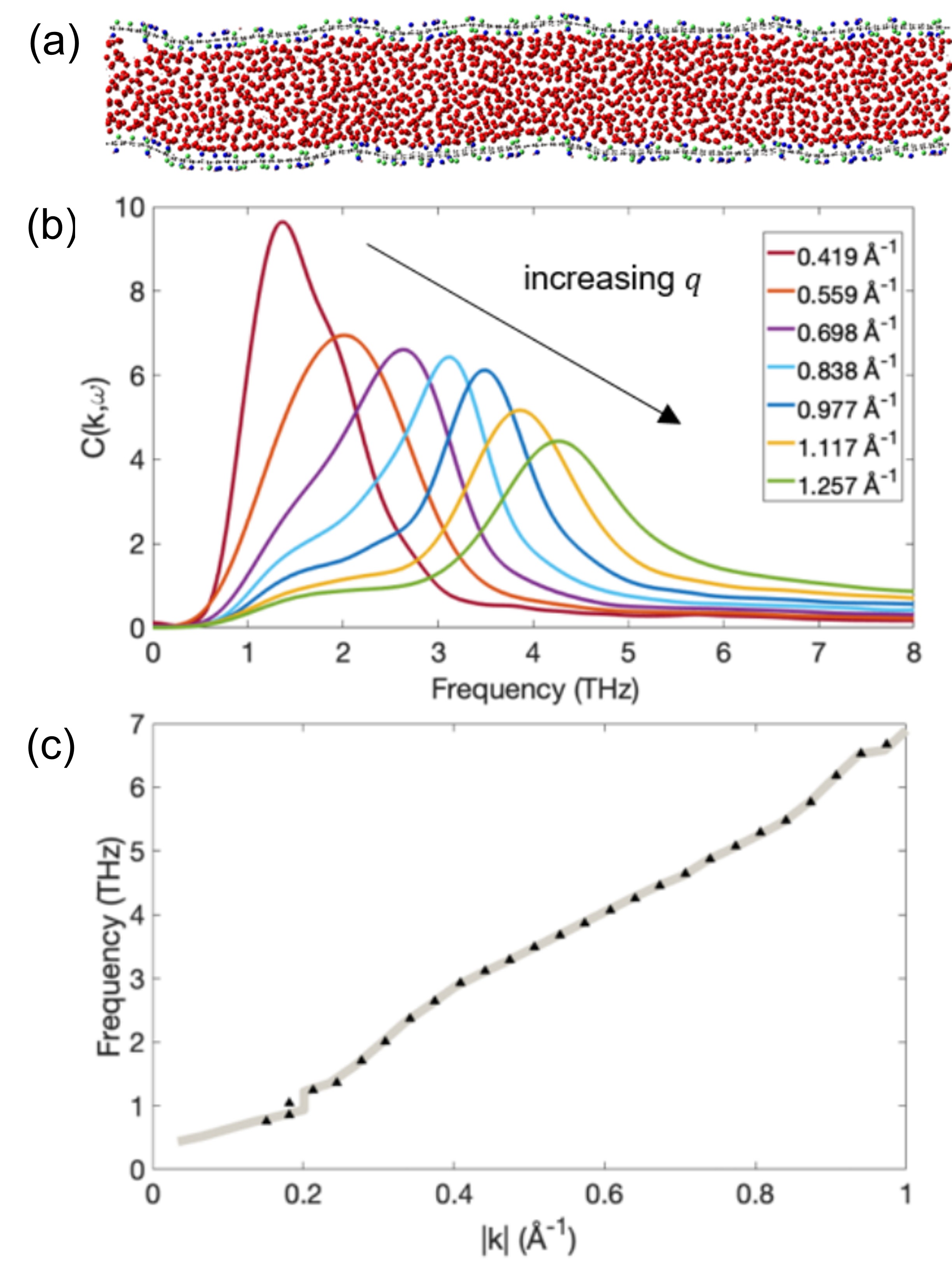}
    \caption{(a) Rendering of an amorphous ice ultra-thin film sandwiched between graphene oxide membranes, from atomistic simulations.
    (b)  The longitudinal current correlations spectrum at different $k_x$ with fixed $k_z = 0.07 \text{\AA}^{-1}$. The arrow indicates the motion of the intensity peak upon moving the wave-vector $k_x$. (c):  The dispersion relation as a function of $|k|$. The gray line represents the computed dispersion relation with wavevector parallel to the xy plane. The black triangles represent the dispersion relation with wavevector $k_z = 0.07 \text{\AA}^{-1}$. Adapted from Ref. \cite{Yu_2022}.}
    \label{fig2}
\end{figure}

This is not what one observes in a "real" nanoconfined system, where the minimum value of $k_{z}$ can be much smaller than $\pi/L$. 
To exemplify this point with data, we report below, in Fig. \ref{fig2}, recently published results from atomistic MD simulations for amorphous ice thin films (Fig. \ref{fig2}(a) shows a rendering of the simulated system). The data show clearly that $k_{z}$ can be much smaller than $\pi/L$: indeed, it can be even $\pi/2L$, i.e. a factor of 2 smaller. 

The ultimate reason for this observation, lies in the unavoidable atomic-scale roughness and irregularities of the interface of the thin film, an effect which becomes ever more important for ultra-thin films just a few nanometers thick or thinner.

\begin{figure}[ht!]
    \centering
    \includegraphics[width=\linewidth]{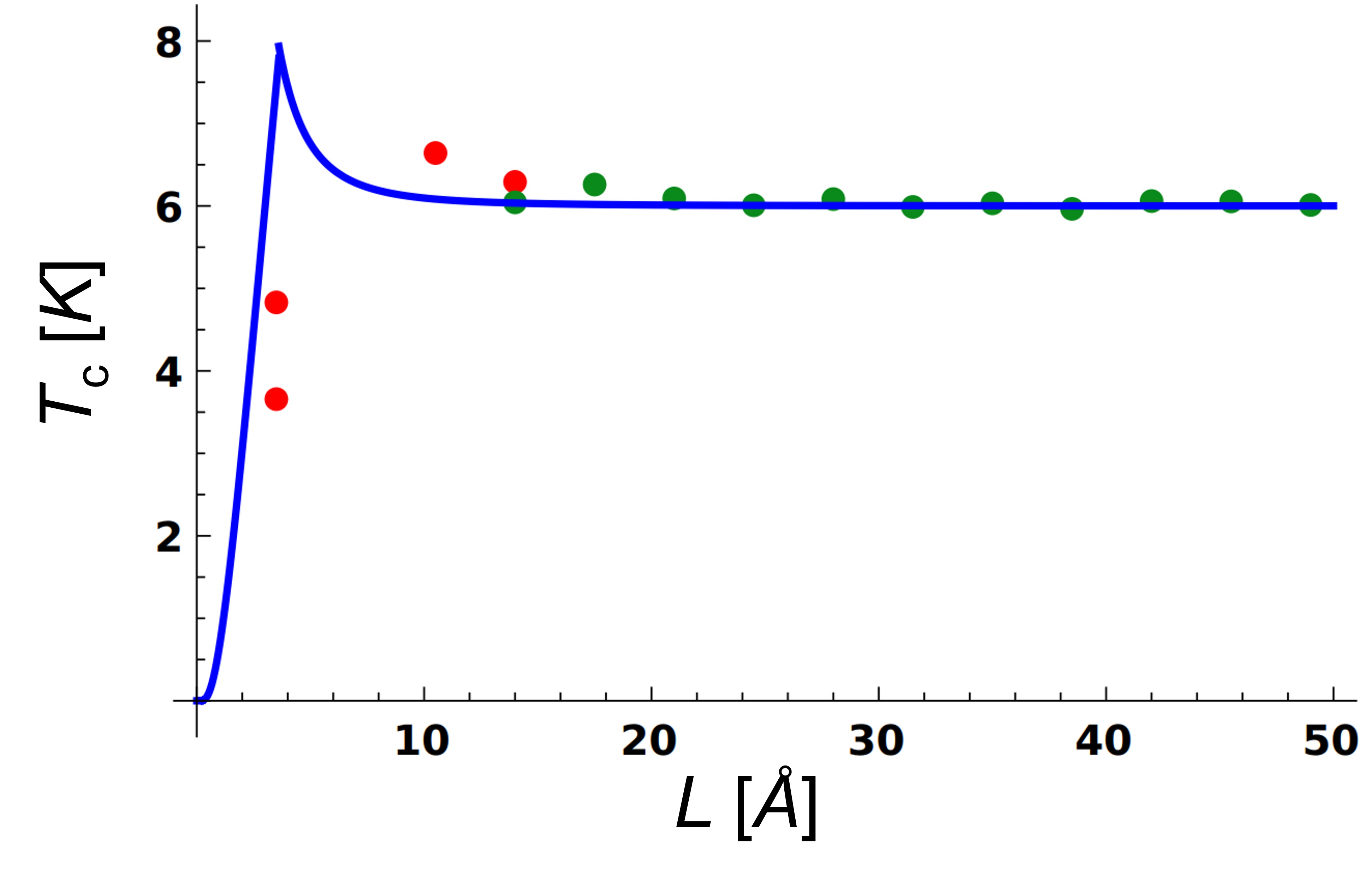}
    \caption{(a) Rendering of a perfectly crystalline copper thin film prepared at 500K, with evidence of strong disorder in the uppermost atomic layer (interface)  \cite{Pavan}. Different colors of atoms in the interface layer correspond to different local crystallographic structures and coordination numbers (different colors correspond to different coordination numbers as explained by the legend). Image courtesy of Massimo Delle Piane.
    (b)  Schematic of an irregular interface of a thin film of thickness $L$ along the confinement dimension (vertical $z$ axis, consistent with Fig. \ref{fig1}(a)). The irregularity of the interface is greatly exaggerated for illustrative purposes.}
    \label{fig3}
\end{figure}

Even for a perfectly crystalline thin film with no defects in the inner atomic layers, the interface presents atomic roughness or, even, significant structural disorder as demonstrated in Ref. \cite{Pavan}. This is illustrated on the example of the interface of crystalline copper, in Fig. \ref{fig3}(a).

With reference to Fig. \ref{fig3}(b), in a real-life thin film with atomic roughness of the interface, the position at which the wavefunction vanishes along the confinement $z$ axis, will be a function of the in-plane coordinates, $x$ and $y$. 
This is a genuine disorder effect, whereby $k_z$ is no longer a good quantum number, since values of $k_z$ will be $n\pi/L(x,y)$ where $L$ is not fixed once and for all, but varies randomly with $x$ and $y$.  Indeed, it is well known from quantum mechanics, that momentum is a good quantum number for hard-wall or periodic BCs only, but not for open BCs. 
As a result, in a real-life thin film there is no discretization of $k_z$ anymore, unlike in an idealized, perfect and smooth rectangular box.
It should also be noted that, for a significantly irregular surface, the two white spheres of hole pockets would also have a corresponding "roughness". However, by taking the average over the roughness, one would retrieve the two perfect white spheres of Fig. \ref{fig3}(b). Hence, the theory evaluated using the two perfect white spheres is correct in an average sense, modulo possible local small fluctuations of the hole pockets envelope around the mean given by the two perfect spheres.

Another consequence of this important fact is the following: if the sample is extended in the $xy$ plane, as it is for thin films, $k$ can still be treated as a quasi-continuous variable \cite{Travaglino_2023,valentinis}, because $|\mathbf{k}|=k=2\pi/\lambda$ obeys the following relation \cite{Kittel,Hill}: 
\begin{equation}
\frac{1}{k^{2}}(k_{x}^{2} + k_{y}^{2}+k_{z}^{2})=1.
\end{equation}

The confinement-induced cutoff on $\lambda$, Eq. \eqref{cutoff}, remains valid, in good approximation, also when the irregularities of the interface are such that $k_z$ is not discretized. The cutoff condition means that a number of quasiparticle states in $k$-space are suppressed due to the confinement along the $z$ direction of the film. 

We shall see what the implications of this cutoff are for the distribution of momentum states of electrons and phonons in the following sections.

\section{Phonons}
Let us reformulate the condition for the cutoff, Eq. \eqref{cutoff}, in terms of the wavevector:
\begin{equation}
k_{min}=2 \pi \cos \theta / L.\label{parametric}
\end{equation}
This is a parametric equation for 
two identical, mirror-image spheres, across the $k_{x}-k_{y}$ plane in a 3D k-space \cite{Phillips}, and is schematically depicted in Figs. \ref{fig1}(b)-(c).
These two (white) spheres, which are contained within the Debye sphere (red), correspond the unoccupied states that cannot be populated due to the confinement. 
As the confinement is further enhanced (or $L$ is decreased), these two white spheres of unoccupied states eventually make contact with the Debye sphere's surface. With additional confinement, the surface corresponding to the highest momentum becomes non-spherical, as shown in
Fig. \ref{fig1}(c). 

\subsection{Vibrational density of states of thin films}
\begin{figure}[ht!]
    \centering
    \includegraphics[width=\linewidth]{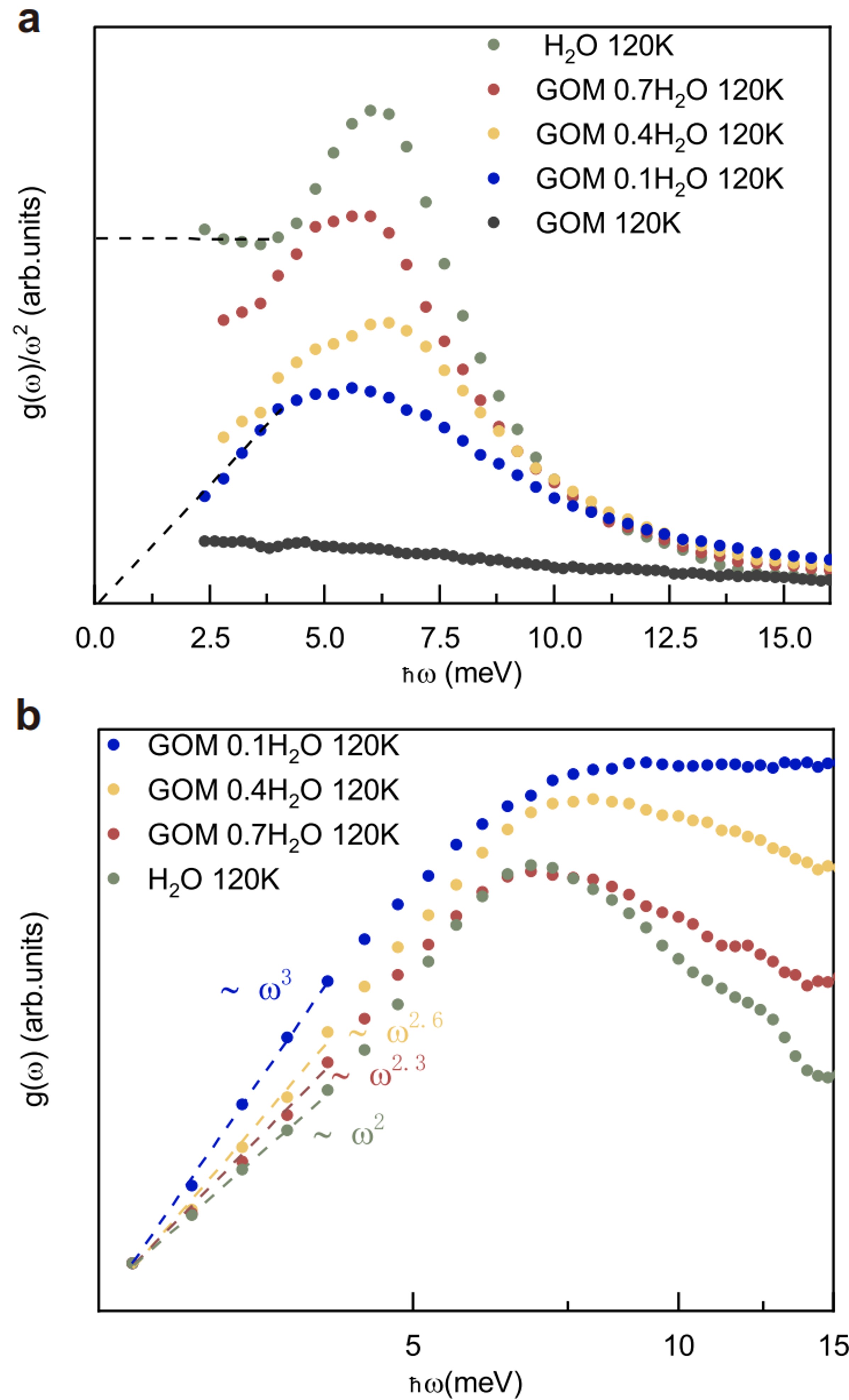}
    \caption{(a) Debye-normalized VDOS of ice thin films (sandwiched between graphene oxide layers) as measured by inelastic neutron scattering, from Ref. \cite{Yu_2022}. Different curves refer to varying thickness $L$ in the range $0.7$ nm to $2$ nm, with $L$ decreasing from top to bottom (the bottom curve is the dry graphene oxide background signal). The top curve exhibits the quadratic Debye law expected for bulk solids, whereas the blue curve exhibits the $\omega^3$ law derived for nanometric thin films.
    (b)  Phonon density of states (not normalized) for a thin film of crystalline ice, of thickness $L$ in the range $0.7$ nm to $2$ nm, with $L$ decreasing from bottom to top,  also from Ref. \cite{Yu_2022}.}
    \label{fig4}
\end{figure}

The occupied volume in k-space can be evaluated exactly using basic solid geometry, as shown in Refs. \cite{Travaglino_2022,Travaglino_2023}, and reads as:
\begin{equation}
Vol_{k}=\frac{L k^{4}}{2}.
\end{equation}
The number of states in k-space with $k<k'$ then readily follows as:
\begin{equation}
N(k<k') = \frac{V}{(2\pi)^{3}}\frac{L k^{4}}{2},
\end{equation}
where $V$ is the total volume of the sample.
The phonon density of states in the thin film then follows immediately, upon defining the speed of sound $v$ such that $\omega = v k$, as:
\begin{equation}
g(\omega) = \frac{d}{d\omega} N(\omega<\omega')=\frac{V}{4\pi^{3}} L  \frac{\omega^{3}}{v^4}\label{VDOS}
\end{equation}
which exhibits a cubic frequency dependence $\sim \omega^{3}$ that was verified both experimentally (inelastic neutron scattering) and by molecular dynamics simulations in Ref. \cite{Yu_2022}. Importantly, the $\omega^3$ law holds for both crystalline thin films as well as for completely amorphous thin films. The above vibrational density of states (VDOS) is for just one phonon polarization, and a factor of three has to be implemented when computing the total internal energy $U$ \cite{Kittel}.
This law is to be contrasted with the standard Debye law for the phonon density of states, exhibiting, instead, a quadratic frequency-dependence:
\begin{equation}
g(\omega) =\frac{V}{2\pi^{2}}  \frac{\omega^{2}}{v^3}\label{Debye}.
\end{equation}
We also note the linear dependence on the film thickness, $L$.
The above $\omega^3$ law for the phonon density of states of thin films has been experimentally verified by inelastic neutron scattering for ultra-thin films of ice of thickness $L\approx 1$ nm, as shown in Fig. \ref{fig4}.

\subsection{Specific heat of thin films}
Since the internal energy of the system can be written as an integral over the VDOS, upon taking the first derivative of the internal energy with respect to the temperature $T$, one readily obtains the following formula for the heat capacity of thin films:
\begin{equation}
C_{v}=120\, \zeta(5)\frac{3}{4\pi^{3}}\frac{L}{v^{4}}k_{B}\left(\frac{k_{B}T}{\hbar}\right)^{4}.\label{key_result}
\end{equation}
This formula exhibits a new $\sim T^{4}$ dependence of the heat capacity on temperature, and a new dependence $\sim L$ on the film thickness. The $\sim T^{4}$ dependence of the specific heat of thin films is radically different from the textbook Debye law $\sim T^{3}$, and is a genuine quantum confinement effect resulting from the cutoff argument Eq. \eqref{cutoff}.

\begin{figure}[h]
\centering
\includegraphics[width=\linewidth]{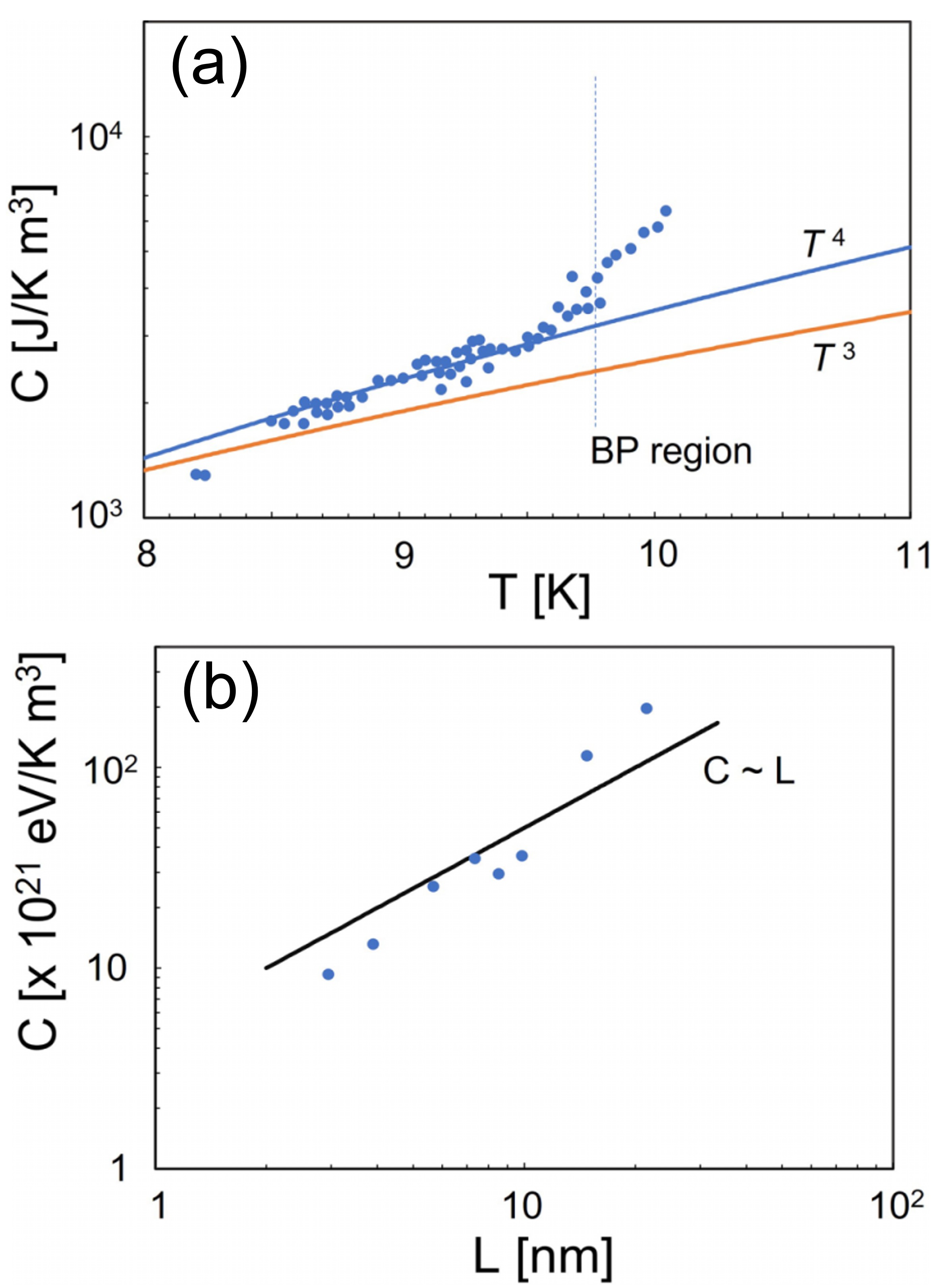}
\caption{(a) Comparison between the $T^4$ temperature dependence of the phonon heat capacity predicted by Eq. \eqref{key_result} (blue line) and experimental data (circles) on NbTiN thin films ($L =6$ nm), adapted from Ref. \cite{Zaccone_heat}. There is only one parameter in the comparison, which is the speed of sound $v \sim 4000$ m/s, here taken as a characteristic value of speed of sound in metals and a plausible value for this material.
The Debye $T^3$ scaling (orange line) is also shown for reference, along with a dashed demarcation line for the onset of the boson peak phenomenon in this class of materials \cite{Zink}. (b) Comparison between the predicted linear scaling of specific heat with thickness $L$ and experimental data for NbTiN thin films, adapted from Ref. \cite{Zaccone_heat}. }
\label{fig5}
\end{figure}

A preliminary comparison of the heat capacity predicted by Eq. \eqref{key_result} with experimental data on NbTiN thin films can be seen in Fig. \ref{fig5}.

These results pave the way for a new understanding of other thermal properties of thin films, including the thermal conductivity. Indeed, the thermal conductivity can be expressed in terms of an integral over the specific heat \cite{Shenshen}, or equivalently, over the VDOS \cite{Larkin,Braun}. In particular, the mechanistic understanding of quantum confinement effects on the phonon statistics, i.e. on the VDOS and on the specific heat, outlined above may represent a solid basis to arrive at a similar mechanistic understanding of the ubiquitously observed increase of thermal conductivity with film thickness \cite{Braun,Nissila,Wang2014}. This also includes the important case of graphene \cite{Balandin,10.1063/5.0244987,Antidormi_2021}, which, in its monolayer state, exhibits an extremely high thermal conductivity \cite{PhysRevB.108.L121412}.

\section{Electrons}

We now consider a metallic thin film, and work in the free-electron approximation. Also in this case, due to the unavoidable roughness of the film surface, the electron wavefunction won't vanish exactly on the surface of an idealized perfectly rectangular box. Hence, due to this and due to the lack of confinement in the $xy$ plane, also in this case, $k_z$ is no longer a good quantum number and we do not expect it to be quantized. 

\subsection{Fermi surface topology of thin films}
The Fermi sphere of a thin film is schematically depicted in Fig. \ref{fig1}(b-c). There are two hole-pocket spheres (symmetric with respect to the origin, along the $k_z$ axis), which represent the unpopulated states in k-space due to confinement. This is because Eq. \eqref{cutoff} applies also in this case.

For a thin film of thickness $L$, this reduction of the available volume for free carriers in momentum ($k$) space is evaluated exactly as (cfr. Fig. \ref{fig1}(b)):
\begin{equation}
    \mathrm{Vol}_{k} = \frac{4}{3} \pi k^3 - 2 \frac{4}{3} \pi \left(\frac{\pi}{L}\right)^3.
\end{equation}

Upon reducing the film thickness further below a threshold $L_c=(2 \pi/n)^{1/3}$, where $n$ is the free carrier concentration, one encounters a topological-type transition first described in depth in Ref. \cite{Travaglino_2023}. At this transition, the Fermi surface undergoes a distortion from the trivial homotopy group $\pi_1(S^2)=0$ of the spherical surface to a surface belonging to a different homotopy group $\mathbb{Z}$, with the new topology depicted in Fig. \ref{fig1}(c). In this situation, the available volume in $k$ space becomes (again see \cite{Travaglino_2023} for a full derivation):
\begin{equation}
    \mathrm{Vol}_k = \frac{4\pi k^3}{3} - V_{inter} = \frac{Lk^4}{2}, 
    \label{case2}
\end{equation}
where $V_{inter}$ denotes the intersection of the two white spheres of hole pockets (states suppressed by confinement) with the original Fermi sphere (Fig. \ref{fig2}).
It could be interesting, in future work, to investigate whether the above topology of the Fermi sea does include two different Fermi surfaces, for holes and electrons, in which case the Bianconi-Perali-Valletta (BPV) theory \cite{BPV1,BPV2} would predict a Fano-Feshbach resonance between two superconducting gaps that could be experimentally checked upon.

\subsection{Electron density of states at varying film thickness}
From this, the electronic density of states $g(\epsilon)$ of free carriers can be easily evaluated \cite{Travaglino_2023}:
\begin{equation}
    g(\epsilon)=\begin{cases}   \frac{V L m^2}{2\pi^3 \hbar ^4 } \epsilon, & \mbox{if }  \epsilon < \frac{2\pi^2 \hbar^2}{mL^2} \\ \frac{ V(2m)^{3/2}}{2\pi^2 \hbar^{3}} \epsilon^{1/2}, & \mbox{if } \epsilon > \frac{2\pi^2 \hbar^2}{mL^2}.
\end{cases}
\label{eq:g}
\end{equation}
with a crossover from the linear-in-energy regime at low energies to the standard Fermi-gas square-root at higher energies.

The crossover is located at an energy $\epsilon^*=\frac{2\pi^2 \hbar^2}{mL^2}$, which depends on the film thickness $L$.

\subsection{Fermi energy as a function of film thickness}
We are now able to derive the Fermi energy $\epsilon_F$ for this system. Since the total number of electrons $N$ in the sample is conserved, we have (always at $T=0$):
\begin{equation}
    N = \int_0 ^{\epsilon_F} g_{s} g(\epsilon) d\epsilon,\label{conserved}
\end{equation}
with the spin-degeneracy factor $g_s =2$.
Upon plugging-in the electron density of states (DOS) of the thin film given by Eq. \eqref{eq:g}, and evaluating the integral in a piecewise fashion, we obtain
\begin{equation}
N=\frac{4}{3} \frac{V(2m)^{3/2}}{(2\pi)^2 \hbar^3} \epsilon_{F}^{3/2} - \frac{4}{3}\frac{\pi V}{L^3},
\end{equation}
and, therefore,
\begin{equation}
    \epsilon_F = \epsilon_F ^{bulk} \left(1+\frac{2}{3} \frac{\pi}{n L^3}\right)^{2/3},
\end{equation}
where $n=N/V$ is the free-carrier density in the sample. One can notice the explicit dependence of the Fermi energy\index{Fermi energy} on the film thickness $L$. The above dependence of Fermi energy on the film thickness recovers expressions that were already proposed in the literature based on the assumption of perfectly smooth rectangular boxes with vanishing hard-wall BCs \cite{ROGERS1987436,BATABYAL2014224}.

Under conditions of strong confinement, it may as well happen that $\epsilon_F < \epsilon^*$. The value of film thickness $L_c$ where this happens depends solely on the free-carrier density $n$ and is given by:
\begin{equation}
L_c \equiv \left(\frac{2\pi}{n}\right)^{1/3}.
\end{equation}
In this regime, we have only one integral since the linear DOS extends up to $\epsilon_F$. 
A simple evaluation gives:
\begin{equation}
\epsilon_F = \frac{\hbar^2}{m}\left[\frac{(2\pi)^3 n}{ L}\right]^{1/2}.
\end{equation}

Upon collecting results for the two regimes, as usual, we can write the Fermi energy as a function of the film thickness across the entire regime of free-carrier density as a piecewise function:
\begin{equation}
    \epsilon_F = \begin{cases} \epsilon_F ^{bulk} \left(1+\frac{2}{3} \frac{\pi}{n L^3}\right)^{2/3} \mbox{ if } L > L_c = \left(\frac{2\pi}{n}\right)^{1/3} \\ \\
     \frac{\hbar^2}{m}\left[\frac{(2\pi)^3n}{ L}\right]^{1/2} \mbox{ if } L < L_c = \left(\frac{2\pi}{n}\right)^{1/3}.\label{Fermi}
    \end{cases}
\end{equation}

\subsection{Resistivity of ultra-thin semiconductor films}
The resistivity of thin films has been a topic of intense research since the advent of modern quantum mechanics and statistical physics, due to its immense technological significance. 
Most of the attention has been traditionally focused on the role of the surface, and, in particular, on the enhanced scattering of free carriers by the interface (in addition to standard scattering by defects and phonons as in the bulk material). In simple terms, thinner films have comparatively more specific interface, such that surface-scattering events will become increasingly more important contributions to the resistivity. As a result, the resistivity increases upon decreasing the film thickness, which is, indeed, what one observes experimentally.
This mechanism lies at the heart of the most widely used theoretical framework for the resistivity and conductivity of thin films, known as the Fuchs-Sondheimer (FS) theory. Originally developed by K. Fuchs in 1938 \cite{Fuchs_1938} and later refined by E. H. Sondheimer \cite{Sond_1952}, the theory is based on an approximate solution to the Boltzmann kinetic equation for the population balance of free carriers, by taking the above mentioned surface-scattering processes into account. 
As reported by Sondheimer \cite{Sond_1952}, simple closed-form expressions for the resistivity contribution of surface scattering are obtained for thick films and thin films, respectively, as $\rho_s/\rho_0=[1+3/(8\kappa)]^{-1}$ and $\rho_s/\rho_0=\{4/[3 \kappa \ln (1/\kappa)]\}^{-1}$, with $\kappa=L/\ell$ where $\ell$ is the mean free path in the bulk material \cite{Sond_1952}. E.g. for silicon, $\ell \approx 20$ nm, hence the crossover between the two formulae occurs around $10$ nm.

The most recent ab-initio calculations have been compared with the predictions of the FS theory in \cite{Yuanyue}. For the case of Cu thin films over a very broad thickness range (spanning from hundreds of nm to about 5 nm), the agreement is not always optimal, in particular for ultra-thin films below 10 nm of thickness \cite{Yuanyue}. While phenomenological approaches based on the full-band model provide a much better fitting of the ab-initio data, still the quantum confinement effects have not been taken into account. 

In the following we propose a combined FS-quantum confinement model which is able to describe the thickness-dependent resistivity of ultra-thin silicon films in a regime where the available experimental data cannot be described by the FS theory alone or by other approaches.

\subsection{Resistivity of c-Si ultra-thin film}
The Fermi level $\mu$ is defined as the Fermi energy $\epsilon_F$ at zero temperature. By thus setting $\epsilon_F \equiv \mu$, for consistency with the semiconductor physics literature, this quantity is given by:
\begin{equation}
    \mu = \begin{cases} \mu_{\infty} \left(1+\frac{2}{3} \frac{\pi}{n L^3}\right)^{2/3} \mbox{ if } L > L_c = \left(\frac{2\pi}{n}\right)^{1/3} \\ \\
     \frac{(2\pi)^{3/2}\hbar^2}{m}\left(\frac{ n}{ L}\right)^{1/2} \mbox{ if } L < L_c = \left(\frac{2\pi}{n}\right)^{1/3}
    \end{cases}\label{chemical}
\end{equation}
where $\mu_{\infty}$ is the Fermi level of the bulk material. 

We consider c-Si semiconductor thin films, which are either intrinsic or weakly-doped such as e.g. the \emph{ex situ}-doped thin films studied recently in Ref. \cite{Duffy}. 
Since, in these materials, there are well-known issues of dopant deactivation, we are going to use the equations for the free-carriers concentration as for intrinsic semiconductors.

For these systems, the concentration of free carriers varies in a broad range, from a lower bound that coincides with the intrinsic material, $n \sim 10^{16}$ m$^{-3}$, to an upper bound of $n \sim 10^{25}$ m$^{-3}$, as noted in \cite{Duffy}.
Here we focus our theoretical analysis on a regime of very weak n-doping where $n \sim 10^{16} - 10^{20}$m$^{-3}$.

In these conditions, $L_c$ is of the order of hundreds of nanometers, and we can thus safely operate in the regime $L<L_c$, using the second of the two relations reported in Eq. \eqref{chemical}. 

In this regime, the concentration of free carriers is given as \cite{Kittel}:
\begin{equation}
    n_i = \sqrt{n_c(T)n_v(T)}\exp{(-E_g/2k_BT)}
\end{equation}
where $n_c(T)=2 (\frac{m_e^*k_BT}{2 \pi \hbar^2})^{3/2}$ and $n_v(T)=2 (\frac{m_h^*k_BT}{2 \pi \hbar^2})^{3/2}$. Here, $m_e^*$ and $m_h^*$ are the effective masses of electrons and holes, respectively, and $E_g$ is the gap energy.
The latter is related to the Fermi level via:
\begin{equation}
    \mu = \frac{1}{2}E_g + \frac{3}{4}k_B T \ln (m_h^*/m_e^*).
\end{equation}
Since the holes are lighter than the electrons, iwe can write:
\begin{equation}
    E_g = 2 \mu - const \cdot k_B T
\end{equation}
where $const >0$. 
This relation reflects the fact that the Fermi level falls exactly in the middle of the energy gap at $T=0$, while it is shifted upwards towards the bottom of the conduction band at room temperature. Because, in the thin film, the Fermi level $\mu$ is a function of the thickness $L$ via Eq. \eqref{chemical}, the above relation implies that the energy gap $E_g$ is a function of $L$. In particular, because the Fermi level increases upon decreasing $L$, the gap energy $E_g$ must increase upon decreasing $L$ (this may no longer be true for monolayer semiconductors where the band structure topology can change significantly, which may lead to a smaller energy gap as demonstrated for PbS monolayers in Ref. \cite{Reviewer1}). Working in the weakly n-doped regime, the conductivity $\sigma$ is given by:
\begin{equation}
    \sigma = (n_i + n_d) e \mu_e 
\end{equation}
where $n_d$ represents the concentration of free carriers due to n-doping, e.g. $n_d \approx [n_c(T) N_d]^{1/2} \exp(-E_d/2k_BT)$, where $N_d$ is the concentration of donors and $E_d$ is the ionization energy of the donor impurity atom. In a first approximation which should remain valid for multi-layer thin films, the ionization energy $E_d$ should not differ substantially from its value in the bulk.
Hence, one can assume the form of $n_d$ to be independent of the film thickness $L$, and that the donor atoms' contribution to the $L$-dependence of the conductivity is negligible. However, this assumption should be treated more carefully upon approaching the $L \rightarrow 0$ limit or the perfect monolayer. That is because, in that case, the electron wavefunction becomes significantly squeezed in the $z$ direction compared to the other directions, and also the total electrostatic potential experienced by the ionizing electron could be significantly different from that in the bulk (this is a phenomenon somewhat analogous to the pressure ionization lowering known in plasma physics \cite{ionization}). Here we shall neglect these effects and leave them for careful consideration in future more microscopically detailed studies.

Furthermore, the mobility $\mu_e = e \tau_e/m_e$ is given in terms of the mean free time between collisions of electrons with phonons and defects, $\tau_e$. Finally, the dependence of the conductivity on the thickness $L$ due to the quantum wave confinement of the electrons is given, in the regime $L < L_c$, by:
\begin{equation}
    \sigma = (n_i + n_d) e \mu_e \sim \exp{(-const/L^{1/2})}. 
\end{equation}
The corresponding resistivity contribution due to confinement, $\rho_c$ is then
\begin{equation}
    \rho_c(L) = 1/\sigma \sim \exp{(const/L^{1/2})}. \label{scaling}
\end{equation}
We now combine this effect of quantum confinement on the conductivity, with the surface-scattering as predicted by the FS theory.
Assuming that Matthiessen's rule is valid, the different effects of quantum confinement and of surface scattering can be summed up as independent contributions \cite{Kittel} to give the total resistivity $\rho(L)$ as:
\begin{equation}
    \rho(L) = \rho_c(L) + \rho_s(L)\label{Matt}
\end{equation}
where $\rho_s(L)$ is given by the FS theory \cite{Sond_1952}.

\begin{figure}[h]
\centering
\includegraphics[width=0.9\linewidth]{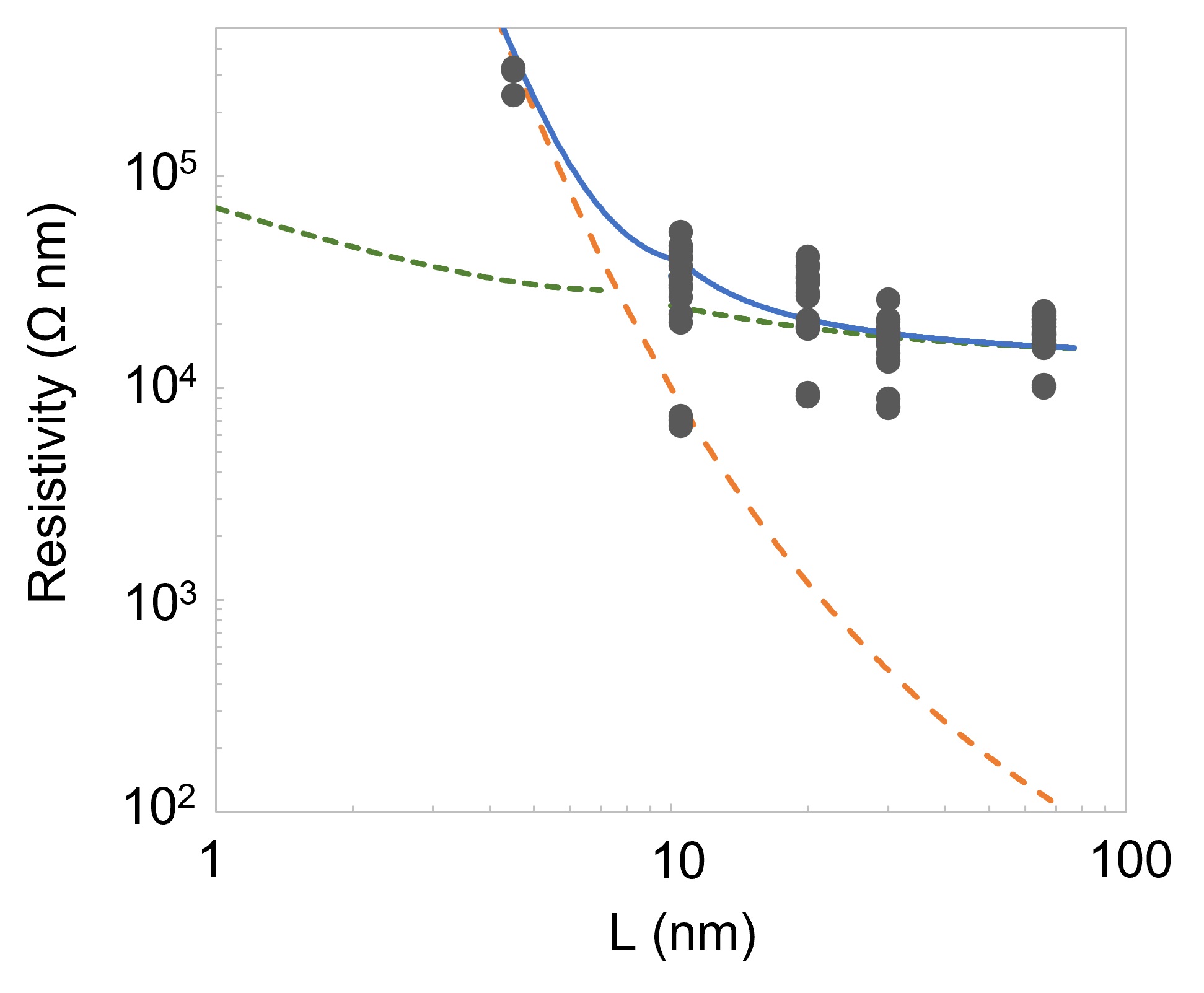}
\caption{Comparison between the theoretical predictions of the proposed model (solid continuous line) obtained by combining the FS surface scattering theory with the electronic confinement model (Eq. \eqref{scaling}) via Matthiessen's rule, Eq. \eqref{Matt}. The green dashed line represents the FS surface scattering prediction (see text for the corresponding equations) without the electron-confinement correction, whereas the orange dashed line represents the electron confinement correction without the FS contribution, given by Eq. \eqref{scaling}. The symbols (circles) are the experimental data from Ref. \cite{Duffy}.  }
\label{fig_duffy}
\end{figure}

Using the form for the confinement-induced resistivity as a function of $L$ given in Eq. \eqref{scaling} for $\rho_c(L)$ and the asymptotic FS expressions mentioned above for $\rho_s(L)$, one thus obtains the fitting of the experimental data of Ref. \cite{Duffy} reported in Fig. \ref{fig_duffy}.

As is clear from Fig. \ref{fig_duffy}, down to $L \approx 10$ nm, the dominant contribution to the resistivity is given by the FS mechanism from surface scattering (green dashed line). As long as the FS is the dominant contribution, the confinement-induced contribution to resistivity is orders of magnitude smaller. At about $L \approx 10$ nm there is a dramatic crossover, with the confinement-induced contribution (orange dashed line in Fig. \ref{fig_duffy}) taking over with respect to the FS contribution. This now becomes the dominant effect as the thickness is reduced below $10$ nm. It is this quantum confinement contribution which is allows the model to capture the sharp increase of resistivity in the range from $10$ nm to $4.5$ nm. Without this contribution, there is no way that the FS theory could describe the experimental data points. This comparison further corroborates the need of including quantum confinement effects, developed by accounting for the presence of surface roughness even in crystalline thin films, when describing the electronic properties of ultra-thin films.
Of course, the caveat always remains that there could be also other, more mundane, contributions to the exponential divergence of resisitivity upon shrinking the thickness to the 2D limit. These include absence of percolation or spatial non-uniformity of activated dopants, which are however impossible to distinguish with the current experimental capabilities as one would need extraordinarily detailed data.

\section{Superconductivity of thin films}
We can now explore the predictions of the theory for ultra-thin metallic superconductors. We shall assume throughout that the conventional electron-phonon pairing mechanism of electrons is valid. Hence, we will first discuss the theory at the level of the Bardeen-Cooper-Schrieffer (BCS) theory \cite{BCS}, for which analytical closed-form expressions of the superconducting critical temperature $T_c$ have been obtained \cite{Travaglino_2023}. Subsequently, we shall implement the quantum confinement model into the more general Eliashberg theory of electron-phonon superconductivity, which allows one to describe also higher levels of electron-phonon coupling. Using the Eliashberg theory, it has been recently possible to obtain a quantitative theoretical description of experimental data of $T_c$ as a function of film thickness for aluminum and for lead thin films \cite{Ummarino_2025}.

\subsection{BCS theory for ultra-thin films}
Let us start with defining
$U_{\Vec{k}\Vec{k'}}$ as the phonon-mediated attractive interaction responsible for the Cooper pairing. In the weak-coupling BCS theory \cite{BCS}, this is simply some negative constant within a Debye-shell beneath the Fermi energy, and zero otherwise:
\begin{equation}
    U_{\Vec{k}\Vec{k'}}= \begin{cases}  -U, & \mbox{if }  |\epsilon-\epsilon_F| < \epsilon_D \\ 0, & \mbox{otherwise }.
    \end{cases}
\end{equation}
Here $\epsilon_D \equiv \hbar \omega_D$ is the Debye energy of the solid, with $\omega_D$ the Debye frequency. 
As usual within BCS theory in its simplest version, the phonons that glue together the Cooper pairs are optical phonons with frequency near $\omega_D$. This fact justifies neglecting the effect of confinement on these phonons (this is because the effects of confinement are mostly affecting acoustic phonons at much lower energy).
Using the Bogoliubov method, we obtain \cite{Cohen}
\begin{equation}
     \Delta_{\vec{k}} = -\sum_{\vec{l}} U_{\vec{k}\vec{l}} \frac{\Delta_{\vec{l}}}{2E_{\vec{l}}}\tanh\left(\frac{\beta E_{\vec{l}}}{2}\right).
\end{equation}
This leads to \cite{Cohen}:
\begin{equation}
    \frac{1}{g(\epsilon_F)U} = \int_0^{\beta_c\epsilon_D/2} \frac{\tanh(x)}{x}dx = \mbox{ln}(1.13\beta_c\epsilon_D),
\end{equation}
where $\beta_c$ indicates the critical value for the Boltzmann factor $\beta=1/k_{B}T$, that is, the value at which the transition from normal metal to superconductor occurs.
Inverting the relation gives \cite{Cohen}:
\begin{equation}
    k_B T_c = 1.13 \epsilon_D \exp\left[-\frac{1}{g(\epsilon_F)U}\right].
\label{eq:t_c_bcs}
\end{equation}
Clearly, the critical temperature $T_c$ depends strongly (in an exponential fashion) on the electron DOS at the Fermi level.
Upon substituting Eq. \eqref{eq:g} for the electron DOS $g(\epsilon)$ of the thin films and we can evaluate this form of the DOS at the thickness-dependent Fermi energy $\epsilon_F$ given by Eq. \eqref{Fermi}. 
We thus obtain:
  \begin{equation}
      T_c = \begin{cases} \frac{4\epsilon_D}{3.52 k_B}  \exp{\left(-\frac{1}{U g^{bulk}(\epsilon_{F}) (1 + \frac{2}{3} \frac{\pi}{n L^3})^{1/3} }\right)} ~~\mbox{ if } L > L_c\\ \\
      \frac{4\epsilon_D}{3.52 K_B}  \exp{\left(-\frac{1}{U g^{bulk}(\epsilon_F)}\frac{(3\pi^2 n)^{1/3}}{\sqrt{2\pi L n}}\right)} ~~~~\mbox{ if } L < L_c.\end{cases}
      \label{eq:t_c}
  \end{equation}
According to this expression, the $T_c$ is a decreasing function of $L$ for $L>L_c$ and an increasing function of $L$ for $L<L_c$. Therefore, this expression predicts a maximum in the $T_c$ as a function of thickness $L$, which occurs exactly at $L=L_c$. This is the thickness value at which the topological-type transition of the Fermi surface occurs, from the Fermi sphere $\pi_1(S^2)=0$ to the distorted Fermi surface with homotopy group\index{homotopy group} $\mathbb{Z}$, cf. Fig. \ref{fig1}(b) and (c). 

The physical origin of this dome of $T_c$ with thickness $L$ lies in the topological distribution of free-electron states in the Fermi sea. For $L>L_c$, cf. Fig. \ref{fig1}(b), as the thickness is reduced, the hole-pocket spheres grow with the consequence that more electron states are pushed to the Fermi surface, which thus increases the electron DOS at the Fermi level, i.e. $g(\epsilon_F)$. In turn, the increase of $g(\epsilon_F)$ within the BCS equation leads to an exponential increase of $T_c$ as $L$ gets reduced.
Conversely, if the thickness is lower than $L_c$, we are now in the situation depicted in Fig.\ref{fig1}(c). Due to the changed topology of the Fermi surface, as the thickness gets further reduced, in this case, the electron states get more and more spread out over a larger Fermi surface. The consequence of this is that, now, $g(\epsilon_F)$ decreases with further decreasing $L$. In turn, this leads to a decreasing trend of $T_c$ as the thickness $L$ is further decreased. 

\subsection{Superconductivity dome with film thickness}
The peculiar trend of $T_c$ as a function of film thickness $L$, with a maximum, has been observed various experimental systems, e.g. in Ref. \cite{doi:10.1126/sciadv.adf5500} for epitaxial aluminum and also recently in Ref. \cite{Dressel} (previous theories also predicted a regime of enhancement of $T_c$ due to confinement \cite{Bianconi}). In all these cases, i.e. both in the experimental data-sets as well as in the predictions of the current theory, there is no visible sign of regular oscillations in the trend of $T_c$ vs $L$, contrary to older theoretical claims \cite{Blatt,PhysRevB.75.014519,doi:10.1126/science.1106675,doi:10.1126/science.1105130}. The reason for this has to be found, again, in the absence of regular discretization for $k_z$ as explained in Section II above.

By using values of the bulk properties, e.g. $g^{bulk}(\epsilon_F)$ and $U$, close to those reported in the literature, a  good agreement between Eq. \eqref{eq:t_c} and experimental data of Pb ultra-thin films from Ref. \cite{lead2} has been demonstrated in Ref. \cite{Travaglino_2023}, supporting the non-monotonic behaviour with the dome in $T_c$. However, Pb is a strong-coupling material with a rather high $T_c$ value and large electron-phonon coupling, which should be more rigorously described by the more general Eliashberg theory. This is discussed in what follows.

\subsection{Eliasbherg theory of superconducting thin films}
For the governing equations of the Eliashberg theory of superconductivity in the Migdal approximation we shall refer the reader to the excellent reviews \cite{revcarbimarsi,ummarinorev}. The standard one-band s-wave Eliashberg equations, when the Migdal theorem holds, can be solved numerically by taking, as the only input, the electron-phonon spectral density $\alpha^2F(\Omega)$ also known as Eliashberg function (where $\Omega$ denotes the phonon frequency). In simple words, this is the phonon frequency-dependent sum over the contributions from scattering processes involving electrons and phonons on the Fermi surface. The Eliashberg function can be either measured experimentally e.g. by tunnelling measurements \cite{Dayan}, or computed via ab-initio methods \cite{revcarbimarsi,GIRI}. 
The other key input to the Eliashberg equations is the electron DOS, which the crucial thickness-dependent quantity, and is given by the quantum confinement theory of Ref. \cite{Travaglino_2023}, Eq. \eqref{eq:g} reported above.
If one removes the approximations of the infinite bandwidth and of taking the electron DOS equal to a constant (i.e. its value at the Fermi level), the Eliashberg equations are slightly more complex and they become four equations \cite{Allen}.
However, when the electron DOS is symmetrical with respect to the Fermi level, the situation is particularly simple because the non-zero self-energy terms are just two, which facilitates the computation.

\begin{figure}[h]
\centering
\includegraphics[width=\linewidth]{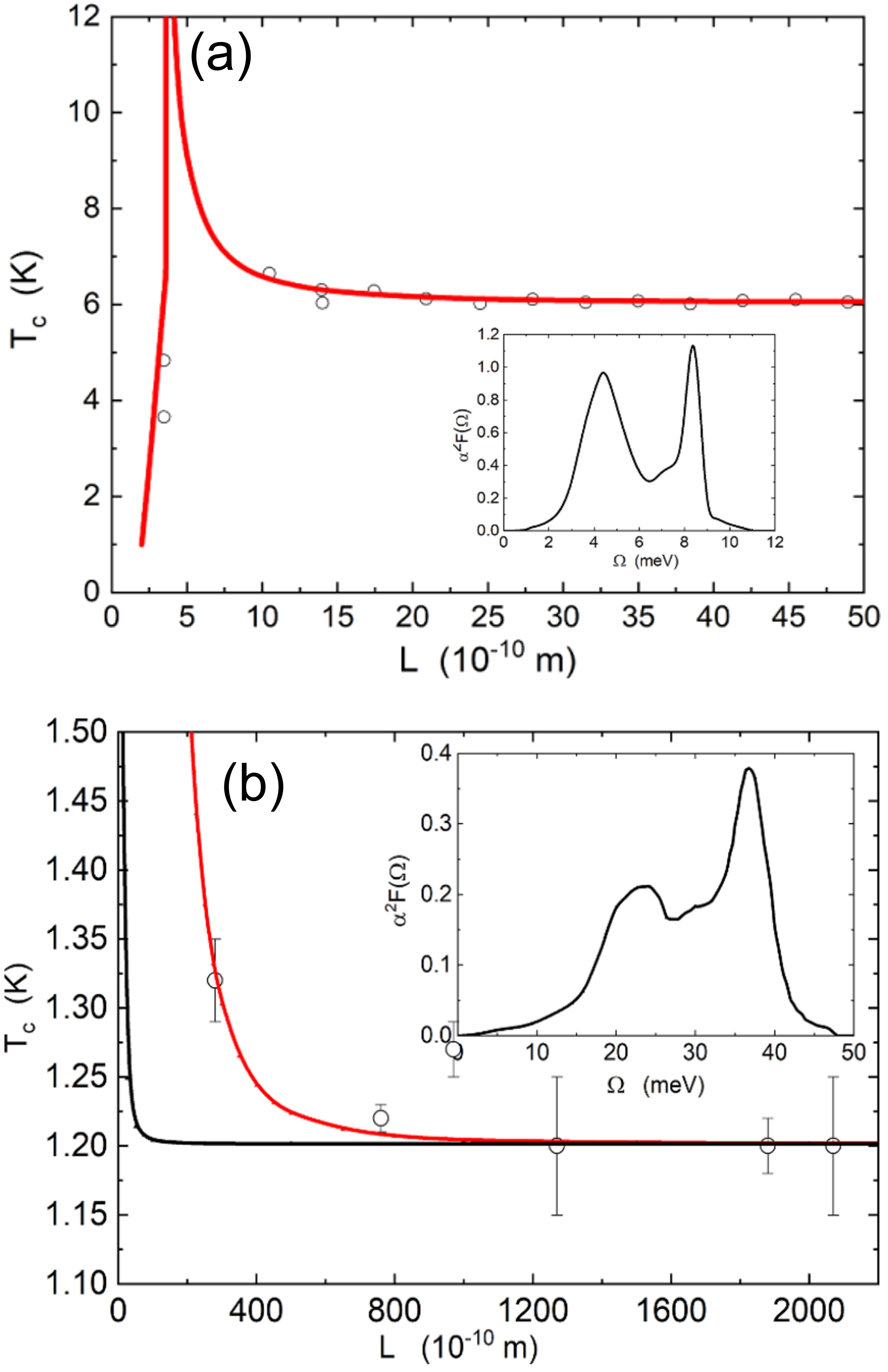}
\caption{(a) Comparison between Eliashberg theory predictions, accounting for quantum confinement, of the critical temperature $T_c$ as a function of film thickness (solid line) and experimental data for Pb thin films (circles). (b)  Comparison between Eliashberg theory predictions, accounting for quantum confinement, of the critical temperature $T_c$ as a function of film thickness (solid line) and experimental data for Al thin films (circles). The experimental data for Pb and Al are taken, respectively, from Ref. \cite{lead2} and Ref. \cite{AlTc}. Adapted from Ref. \cite{Ummarino_2025}. The insets show the Eliashberg spectral function $\alpha^2F(\Omega)$ for the two materials.}
\label{fig6}
\end{figure}

This is the scheme used to obtain a quantitative theoretical prediction of the critical temperature $T_c$ as a function of thickness in excellent agreement with experimental data for Al and Pb thin films in Ref. \cite{Ummarino_2025}, shown in here in Fig.\ref{fig6}.

\subsection{Non-superconducting elements become superconductors near the 2D limit}
\begin{figure}[h]
\centering
\includegraphics[width=\linewidth]{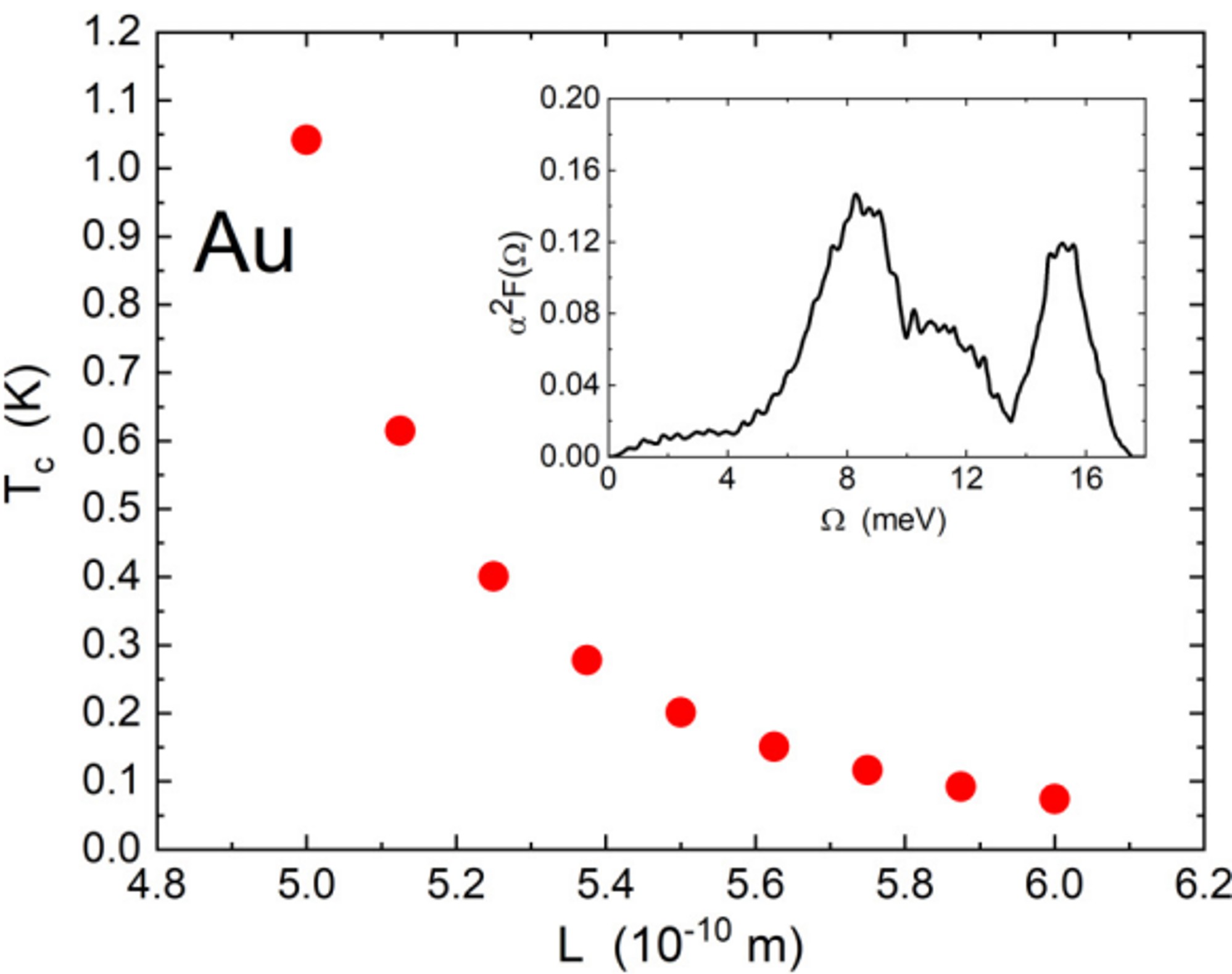}
\caption{Eliashberg theory predictions, accounting for quantum confinement, of the critical temperature $T_c$ as a function of film thickness $L$ for gold thin films. The inset shows the Eliashberg spectral function $\alpha^2F(\Omega)$ as computed via ab-initio methods \cite{GIRI}. Adapted from Ref. \cite{gold}.}
\label{fig7}
\end{figure}

We learn in high-school that noble metals, such as gold and silver, are excellent conductors of heat and electric current. However, they are not superconductors, or at least their critical superconducting temperature $T_c$ is too low to be measured with standard equipment. We have seen, however, that quantum confinement, in the regime $L>L_c$ can strongly enhance the superconducting $T_c$ of a given material, due to the growing hole-pockets which push more states to move to the Fermi surface. We have also seen that for good conductors, i.e. for materials with a large concentration of free carriers, $n$, the maximum at $L_c$ is pushed to extremely low values of thickness, basically to the 2D limit. 
This implies that one can take advantage of the confinement-induced enhancement of $T_c$ basically down to the 2D limit, for good conductors. 
This physical consideration is behind the idea of exploring the behaviour of $T_c$ for ultra-thin noble metal films, near the 2D limit.
This was done, again by means of Eliashberg theory implementing the quantum confinement model described above and the most accurate ab-initio calculations of the Eliashberg function for noble metals. The results, published in Ref. \cite{gold}, are somewhat surprising, as the reveal the possibility that ultra-thin films of gold, about 0.5 nm, may be superconductors with the same critical temperature of aluminum (the latter is the most used material for qubits). The results are shown in Fig. \ref{fig7}.

\begin{figure}[h]
\centering
\includegraphics[width=\linewidth]{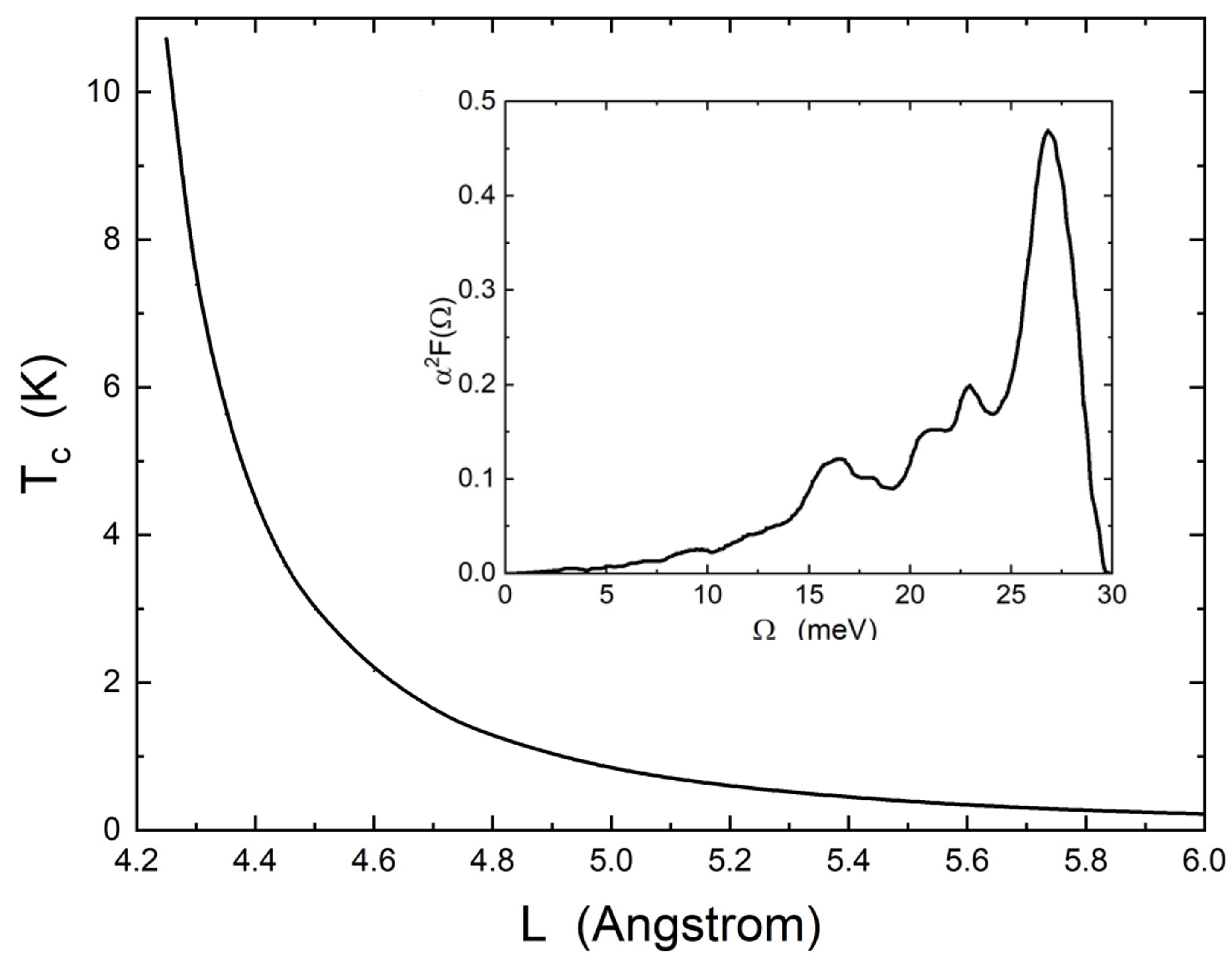}
\caption{Eliashberg theory predictions, accounting for quantum confinement, of the critical temperature $T_c$ as a function of film thickness $L$ for magnesium (Mg) thin films. The inset shows the Eliashberg spectral function $\alpha^2F(\Omega)$ as computed via ab-initio methods \cite{chulkov}. Adapted from Ref. \cite{magnesium}.}
\label{fig8}
\end{figure}

Finally, even magnesium (Mg), an alkaline-earth metal that is lighter than aluminum, is well known to be a good conductor but not to be a superconductor. Even in this case, however, quantum confinement leads to such increase in the DOS at the Fermi level, that Mg becomes an excellent superconductor when cast into ultra-thin sub-nanometer sheets. The same type of calculation \cite{magnesium}, using Eliashberg theory and the Eliashberg function computed from ab-initio simulations, predicts that Mg can achieve a superconducting critical temperature as high as 10 K when the thickness is about 0.4 nm, as shown in Fig. \ref{fig8}. Also in this case, there are no adjustable parameters in the prediction. If confirmed experimentally, this would be a particularly attractive discovery for many technological applications (e.g. quantum computing, quantum electronics), because Mg would be a superconductor at temperatures well above the liquid helium boiling temperature. 

Objectively, an experimental verification of this prediction is difficult, because the effect may easily be obscured, e.g. by proximity effects. Also, ultra-thin films of gold may be extremely brittle and mechanically unstable.
However, good candidate systems to detect this effect have been recently developed experimentally, which include: macroscopically large, nearly freestanding 2D gold monolayers, consisting of nanostructured patches formed on on an Ir(111) substrate and embedding boron (B) atoms at the Au/Ir interface \cite{Alexei}.
Another candidate experimental system could be the ultra-thin nanowires obtained via nano-molding \cite{Schroers}, which, in future technical improvements, may reach the atomic-scale thickness. 

\subsection{Tuning the thin film superconductivity with electric fields}
Recent experimental work has revealed that superconductivity in thin metallic films can be suppressed by applying a strong enough external DC electric field, denoted as $E_{cr}$ \cite{Giazotto1,Giazotto2,review}. Experimental evidence has shown that external electric fields (EF) on the order of $\sim 10^8$ V/m are required to suppress the supercurrent in metallic thin films with a thickness of around 20 nm \cite{review}. 
These findings are of great technological importance, because supercurrent field-effect transistors have huge potential for future classical \cite{Mukhanov04,Tolpygo16} and quantum computation \cite{Gambetta17} nanodevices. 

In spite of this intense experimental activity, the microscopic mechanism by which an external electric field (EF) is able to penetrate a superconductor on a length-scale much larger than the Thomas-Fermi length of the normal state, has remained unclear. Numerical and experimental evidence has shown that the penetration length of an external static electric field into Niobium-based thin films can be as large as 4 nm, hence closer to the London depth than to the Thomas-Fermi length \cite{Gonnelli,Piatti}. Some electrodynamic covariant theories \cite{Tajmar,Hirsch}, building on an original intuition of F. London, seem to justify that an external EF penetrates into the SC phase on a length scale comparable to the London length.

Whatever the exact penetration depth of the EF into the superconductor, standard electrodynamics shows that the EF amplitude decays exponentially from the interface into the thin film, with a characteristic decay length. If the decay length is a few nanometers, the EF will be non-zero also beyond the the decay length, because an exponentially-decaying function is identically zero only at infinite distance.
This, in turn, will lead to a finite probability of Cooper pair breakage via tunnelling enabled by the EF. A microscopic theory of this effect, within a simplified version of the Eliashberg theory, has been derived recently in Ref. \cite{Fomin}. The theory predicts the critical value of EF needed to suppress superconductivity in metallic thin films, as a function of the film thickness $L$.

The problem of splitting a Cooper pair by an electric field is analogous to the textbook problem of electric field-induced dissociation of an s-wave bound state. This is because, within BCS theory, a Cooper pair is described by a s-wave bound state satisfying the Schr\"{o}dinger equation for two electrons interacting via an effective attractive force \cite{Cooper,BCS}, with a real-space description originally suggested by Weisskopf \cite{Weisskopf}. 
The solution for the bound-state dissociation under an electric field is well known \cite{Landau}, and has been used to describe the Cooper pair splitting by an external EF in 
\cite{Fomin}. The formula for the critical EF magnitude needed to split the Cooper pair is given by \cite{Fomin}:
\begin{equation}
E_{cr}=\frac{2 \Delta}{e\,\xi}, \label{crit}
\end{equation}
where $\Delta$ is the BCS energy gap, $e$ is the electron charge, and $\xi$ is the coherence length, which is obtained by the solution of Eliashberg equations \cite{revcarbimarsi}.

The above formula for $E_{cr}$ can be derived by considering the Schr\"{o}dinger equation for an electron initially bound in a s-wave bound state (the Cooper pair) of energy depth $\Delta$, and subjected to an external electric field of magnitude $E = |- \nabla V|$:
\begin{equation}
    \left(\frac{1}{2}\nabla^{2} + \mathcal{E}+u(r) - E z\right) \psi = 0 \label{Landau}
\end{equation}
where $\mathcal{E}$ is the energy, $E$ is the electric field magnitude, $z$ is the spatial coordinate along which the EF is pointing, and $\psi$ is the wavefunction. In the above equation, atomic units have been used. Furthermore, the attractive potential is schematically given by a spherical well: $u(r)=-\Delta$ for $0 \leq r \leq \xi$ and zero otherwise, where $\Delta$ is the BCS energy gap and $\xi$ is the coherence length \cite{Weisskopf}. The solutions to Eq. \eqref{Landau} are obtained by separation of variables in parabolic coordinates, and can be found in textbooks such as in Ref. \cite{Landau}. From the solution to Eq. \eqref{Landau}, one obtains the characteristic critical field $E_{cr}$ to break the Cooper pair as follows.

While the s-wave bound state (the Cooper pair) is spherically symmetric, the electric field is directed along a certain spatial direction, which could be any direction in the solid angle. Hence, as shown with full details in the textbooks, pp. 296-297 of \cite{Landau}, one solves the 
Schr\"{o}dinger equation Eq. \eqref{Landau} in parabolic coordinates, and uses the solution to compute the probability current of the electron escaping away from the bound state in the direction of the EF (i.e. the coordinate $z$ in Eq. \eqref{Landau}). The result for the probability $w$ of the electron tunnelling away from the bound state, in atomic units, is \cite{Landau}:
\begin{equation}
    w \sim \exp\left(-\frac{2}{3 E}\right)
\end{equation}
where $E$ is the magnitude (absolute value) of the electric field. For a s-wave bound state of unitary depth energy and unitary radius, converting from atomic to physical units, the above formula from \cite{Landau} reads as:
\begin{equation}
    w \sim \exp\left(-\frac{2}{3}\frac{E_a} {E}\right) \label{esc}
\end{equation}
where 
\begin{equation}
    E_a = \frac{2 R_H}{e a_0},
\end{equation}
with $R_H$ the Rydberg energy and $a_0$ the Bohr radius (both are equal to unity in the atomic units used in Landau's derivation \cite{Landau}).
Hence the critical electric field to dissociate the bound state is:
\begin{equation}
    E_{cr}=\frac{2 R_H}{e a_0}.
\end{equation}

For a generic s-wave bound state of depth energy $\Delta$ and radius $\xi$, the critical EF needed to dissociate the bound state is thus \cite{Fomin}
\begin{equation}
E_{cr}=\frac{2 \Delta}{e \xi} \nonumber
\end{equation}
which is just the above formula Eq. \eqref{crit}.

In the above equation for the critical EF, a possibly material-dependent parameter is the coherence length $\xi$. The latter is given by \cite{deGennes}:
\begin{equation}
    \frac{1}{\xi}=\frac{1}{\xi_0}+\frac{1}{\ell},
\end{equation}
where $\xi_0$ is the intrinsic (Pippard) coherence length, and $\ell$ is the mean free path. Thin films, such as those used in the supercurrent field effect devices, have a microstructure characterized by microcrystallites, the size of which sets the value of $\ell$. Since, typically, $\ell \ll \xi_0$ (because $\xi_0$ can be tens or hundreds of nanometers), the coherence length $\xi$ is controlled by $\ell$, and, hence, by the disorder, and $\xi \approx \ell$.
Because for experimental metallic thin film systems the disorder is always present \cite{Sidorova}, in the form of small grains (crystallites) that are randomly packed, we have $\xi \approx \ell$, and, therefore:
\begin{equation}
E_{cr}=\frac{2 \Delta}{e\,\ell}. 
\end{equation}
Being in the diffusive regime where the coherence length $\ell$ has a small value is implemented in the Eliashberg spectral function, which is not that of a bulk superconductor but that of a thin film.
Knowing the gap energy $\Delta$ for a given material from the Eliashberg theory, one can estimate the critical electric field $E_{cr}$ for superconductivity suppression inside the film. 

For the example of NbN, the values of the critical electric field needed to suppress the superconductivity in 10-30 nm-thick thin films are of the order of $10^7$ V/m, under the assumption of no-screening. This estimate is one order of magnitude lower than the experimental values of the order of $10^8$ V/m reported in the literature for films of comparable thickness \cite{review}. 

As already mentioned, this estimate still assumes a perfect penetration of the EF inside the sample, or, in other words, does not account for the screening of the EF inside the sample. To compute the magnitude of the external EF that has to be supplied to cause the suppression of superconductivity, the screening effects need to be taken into account. It is easy to show that, with a penetration depth of the EF of about 4 nm \cite{Gonnelli,Piatti}, the predicted critical value of EF for the suppression of superconductivity becomes of the order of $10^8$ V/m, in agreement with experimental measurements \cite{review}.

\section{Conclusions and outlook}
We attempted to provide a holistic view of the effects that quantum confinement has on the physical properties of thin films, with a special attention to the case of ultra-thin films with a thickness lower than 10 nm. The starting point is the basic physics of quantum wave propagation through a slender rectangular box, with confinement along the vertical direction, and no confinement in the orthogonal plane. The confinement imposes a cut-off on the wavelength of the quasiparticles states that can populate the sample, which leads to simple mathematical forms of the corresponding density of states and k-space topology. 
Importantly, due to the unavoidable disorder and non-smoothness of the interfacial atomic layers at the surface of the film, there is no discretization of the wavevector $k_z$ along the confinement direction in real-life thin films, because standard hard-wall boundary conditions do not apply ($k_z$ is no longer a good quantum number). This fact cannot be captured by traditional theoretical approaches based on an ideal perfectly smooth rectangular box, but is, instead, well-captured by the new confinement approach reviewed here.

In particular, the thin-film confinement leads to a $\omega^3$ form of the phonon DOS, instead of the $\omega^2$ Debye law, a theoretical prediction that has been confirmed experimentally by inelastic neutron scattering for ultra-thin films of ice \cite{Yu_2022}, with a gradual crossover from $\omega^3$ to $\omega^2$ as the thickness increases, as also confirmed by MD simulations. The $\omega^3$ law for the phonon DOS leads to a $T^4$ law for the heat capacity of ultra-thin films at low temperature, instead of the Debye $T^3$ law. In future work, it will be of great interest to use these results for a quantitative theory of thermal conductivity in ultra-thin and quasi-2D materials, such as graphene, van der Waals materials and layered films. 
Applying the same analysis of confinement to free electrons in thin metallic films, leads to a simple form of the electron DOS, which features a linear-in-energy trend at low-energy, which then crosses over into the standard square-root behaviour at a characteristic energy that depends on the film thickness and on the free-electron density. The theory also shows how two spherical hole-pockets of forbidden states grow inside the Fermi sphere as the thickness is decreased, up to the point where the spherical Fermi surface transitions into a surface with a different homotopy group $\mathcal{Z}$. This transition coincides with the transition between the two regimes in the electron DOS and with a change of confinement-controlled redistribution of momentum states on the Fermi surface. For thickness $L>L_c$ the growing hole-pockets inside the Fermi sphere push more states towards the Fermi surface, thus increasing the DOS at the Fermi level. Instead, for $L<L_c$, the new $\mathcal{Z}$ surface becomes more extended as $L$ keeps decreasing, and therefore the momentum states at the surface become more spread out, implying that the electron DOS at the Fermi level decreases. 
This mechanism has fundamental implications for the electronic conduction in ultra-thin films, such as c-Si films with thickness below 10 nm, which fall in the regime $L<L_c$. We have demonstrated that this mechanism of "dilution" of electron states at the Fermi level of the $\mathcal{Z}$ surface leads to an exponential increase of the resistivity upon decreasing the thickness. This effect is superimposed on the Fuchs-Sondheimer (FS) surface-scattering contribution and is responsible for the exponential increase of resistivity observed experimentally in \cite{Duffy}, which cannot be explained or reproduced by the FS theory.
The same quantum confinement mechanism is responsible for the dome in the critical temperature for the normal metal-to-superconductor transition as a function of film thickness, observed experimentally in several systems. Again, the maximum in $T_c$ coincides with the critical thickness $L_c$ at which the Fermi surface distortion induced by confinement takes place. 
Implementation of the the electron DOS as a function of thickness into the BCS theory leads to analytical solutions for the $T_c$ as a function of thickness. For a fully quantitative comparison, implementing the confinement model in the Eliashberg theory of superconductivity has recently led to the first quantitative theoretical description of the $T_c$ as a function of thickness for two experimental systems, i.e. aluminum and lead thin films \cite{Ummarino_2025}, in excellent quantitative agreement with the available experimental data. The same scheme led to the surprising prediction that few atomic layers of gold can become superconducting with the same $T_c$ of aluminum \cite{gold}.
Finally, the same confinement model can explain the existence of a critical value of electric field to suppress superconductivity in thin films \cite{Fomin}, a fact experimentally demonstrated in seminal experiments by F. Giazotto and co-workers \cite{Giazotto1}.

This holistic framework, which describes electrons, phonons and superconductivity in thin films, can be extended in future work in several directions:\\
(i) mechanistic understanding of thickness effect on thermal conductivity of ultra-thin and quasi-2D materials \cite{Balandin};\\
(ii) extension of the framework to include effects of lattice anharmonicity \cite{Setty_2024}, distortions \cite{Jiang_2023}, and structural disorder \cite{amorphous} in the interior of the film (not just the interface);\\
(iii) application of the framework to high-temperature cuprates superconductors \cite{Jiang_2024}, for which atomically-thin film are now available experimentally \cite{Poccia};\\
(iv) improve the understanding of dielectric properties of thin films as a function of thickness, from ferroelectric \cite{Zaccone_perm} to superconductors \cite{Hirsch};\\
(v) extend the framework to different shapes of 3D nanostructures such as nanowires \cite{wire,Cuniberti,Joyce}, nano-rings \cite{Fomin_book} (for the latter, see Ref. \cite{Landro}) and M{\"o}bius strips \cite{Wang2023}, quantum-dots supercrystal assemblies \cite{Schall}, 2D random networks of atomic clusters and nanoparticles \cite{Milani}, and artificial heterostructures of alternating superconductor and normal layers \cite{Perali_2024}.

\section*{Data Availability}
All data that support the findings of this study are included
within the article (and any supplementary files).

\subsection*{Acknowledgments} 
Dr. Ray Duffy (University College Cork and Tyndall National Institute, IE) is gratefully acknowledged for providing the original data set from Ref. \cite{Duffy}.
The author gratefully acknowledges funding from the European Union through Horizon Europe ERC Grant number: 101043968 ``Multimech'', from US Army Research Office through contract nr.   W911NF-22-2-0256, and from the Nieders{\"a}chsische Akademie der Wissenschaften zu G{\"o}ttingen in the frame of the Gauss Professorship program. 

\bibliography{refs}

\end{document}